\documentclass[10pt,twocolumn]{article}

\usepackage[utf8]{inputenc}
\usepackage[T1]{fontenc}

\usepackage{graphicx}

\usepackage{lipsum}
\usepackage{xcolor}

\usepackage{authblk}
\usepackage[margin=2cm]{geometry}
\usepackage{mathtools, cuted}

\usepackage{url}
\usepackage{bm}
\usepackage{amsfonts}
\usepackage{amsmath}
\usepackage{amssymb}
\usepackage{mathrsfs} 
\usepackage{siunitx}
\usepackage[numbers]{natbib}

\title{Lifetime of vertical giant soap films: role of the relative humidity and film dimensions}
\author[1]{Marina Pasquet}
\author[1]{François Boulogne}
\author[1]{Frédéric Restagno}
\author[1]{Emmanuelle Rio$^{\ast}$}
\affil[1]{Université Paris-Saclay, CNRS, Laboratoire de Physique des Solides, 91405, Orsay, France.}

\date{\today}
\begin{document}

\twocolumn[
    \begin{@twocolumnfalse}
        \maketitle
        \begin{abstract}
We consider the lifetime of rectangular vertical soap films and we explore the influence of relative humidity and both dimensions on the stability of large soap films, reaching heights of up to 1.2 m. Using an automated rupture detection system, we achieve a robust statistical measurement of their lifetimes and we also measure the film thinning dynamics. We demonstrate that drainage has a negligible impact on the film stability as opposed to evaporation. To do so, we compare the measured lifetimes with predictions from the Boulogne \& Dollet model \cite{BoulogneDollet2018}, originally designed to describe the convective evaporation of hydrogels. Interestingly, we show that this model can predict a maximum film lifetime for all sizes.        \end{abstract}
    \end{@twocolumnfalse}
]

\section{Introduction}

In their performances, bubble artists daily make soap films of large sizes, as evidenced by some world records such as the highest indoor bubble exceeding 10 meters high \cite{WorldRecordGraeme} or the largest outdoor and indoor bubbles exceeding respectively 96 \cite{WorldRecordGary} and 21 \cite{WorldRecordGaryindoor}~m$^3$ in volume.
The reported lifetimes of these giant objects are generally of the order of a few seconds, but they depend strongly on the composition of the soap solution \cite{pasquet2022optimized}, as well as on the ambient air humidity. 

Large soap films are also the subject of scientific studies \cite{Ballet2006, Cohen2017,mariot2021new, Rutgers2001, Kellay2002,pasquet2023thickness}. 
Experiments have been conducted on the impact of air humidity on the stability of foams \cite{Li2010,Li2012}  and on centimeter scale soapy objects, bubbles and films \cite{Champougny2018, miguet2020stability, Poulain2018PRL}.
They demonstrated the crucial role of humidity on the lifetime of soapy objects, and some studies enlighten the complex role of the nature of surfactants on evaporation \cite{la_mer_evaporation_1965,poulain2018ageing, Pasquet2022}. 

To date, the thinning rate of surface bubbles is actually better understood than the one of flat soap films, the rotational symmetry being helpful for a quantitative description.
More specifically, the thinning rate of the soap film surrounding surface bubbles is influenced by drainage, and also by evaporation when the atmosphere is not saturated in humidity. 
The question of the rupture thickness and the nucleation mechanism of the holes in soap films is still widely open, nevertheless it appears that it only widens the distribution of lifetime around an average value fixed by the thinning rate. 
Within this frame, the lifetime of surface bubbles can be predicted by describing their thinning through drainage and evaporation \cite{miguet2020stability,Poulain2018PRL}. 
In particular, the description of evaporation must take into account the natural air convection \cite{miguet2020stability}. 

The drainage of both surface bubbles and vertical flat films has been proven to be due to marginal regeneration \cite{Mysels1959,Seiwert2017,miguet2021marginal}:
in the vicinity of the meniscus, a groove develops \cite{Aradian2001}, which destabilizes in thin patches whose smaller apparent density \cite{adami2014capillary} makes them rise.
Since the interface can be considered as incompressible \cite{Seiwert2017}, the consecutive \textit{sliding puzzle dynamics} \cite{gros2021marginal} with rising thin patches and descending thick zones is at the origin of the drainage \cite{Mysels1959,Seiwert2017,miguet2021marginal}.

For centimetric vertical flat films, the exact thinning law still lacks a quantitative description.
In this geometry, the marginal regeneration indeed occurs not only at the bottom meniscus but also on the vertical borders. 
Two gutters, close to the lateral menisci, allow a rapid rise of thin areas, resulting from the destabilization of the lateral groove \cite{Mysels1959,Seiwert2017}.
This contribution to drainage is not fully described in the literature. 
Nevertheless, a previous study \cite{Champougny2018} has shown experimentally that for thicknesses smaller than a few hundreds of nanometers, the evaporation becomes predominant in the thinning rate. Concerning the lifetime of flat films, Frazier \textit{et al} \cite{Frazier2020} propose a measure for films of a few centimeters high for solution containing a mixture of dish washing liquid and lubricant  as a function of humidity and lubricant concentration. They show that the addition of lubricant increases the lifetime. Furthermore, they provide a measure of the film lifetime with the atmospheric humidity, which increases drastically above 80\%.

Recently, we developed a setup allowing to study metric films, which can be qualified as giant soap films \cite{mariot2021new}. 
We measured their thickness profile under gravity and described their dynamics during their generation \cite{pasquet2023thickness}.
In this paper, we use the same setup and measure the evolution of the lifetime of giant films with ambient humidity and with their dimensions.
We show that the results are in agreement with a description, in which evaporation has a primary role in thinning, allowing to understand  the maximal lifetime for the films.

\section{Experimental setup}

\begin{figure*} 
  \centering
    \includegraphics[width=\linewidth]{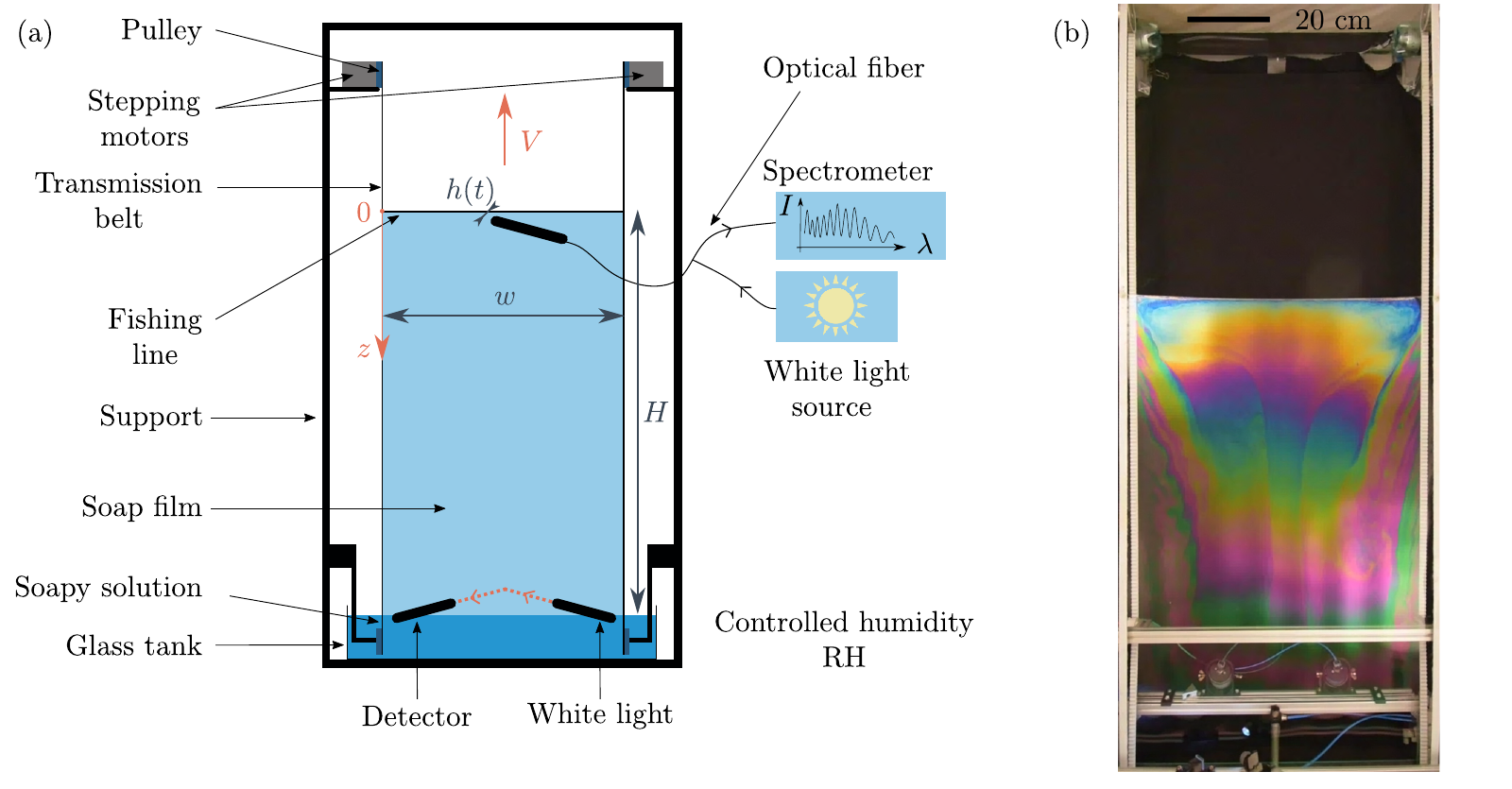}
  \caption{(a) Schematic representation of the experimental setup allowing generation and automated lifetime measurements of giant soap films, with the notations used in the article. A spectrometer measures the film thickness at a given position. (b) Photograph of a static soap film of height $H = 120$~cm and width $w = 70$~cm, made with a solution of Fairy diluted to 4 \%. The spectrometer is removed for the purpose of illustration.}
  \label{fig:Setup}
\end{figure*} 

To study the lifetime of giant soap films, we have used the experimental setup presented in Fig. \ref{fig:Setup} (a) and a photograph of a studied film is proposed in Fig. \ref{fig:Setup} (b). In this section, we give the main characteristics of this setup for which more details are available in a dedicated article by Mariot \textit{et al.} \cite{mariot2021new}. 

The films are generated by pulling a horizontal wire  out of a glass tank.
For each experiment, the tank is filled with 5 liters of solution made of 4 wt.\% dishwashing liquid (Fairy from Procter \& Gamble, with an anionic surfactant concentration between 15 and 30 \%) with ultrapure water (resistivity greater than 18.2 M$\Omega \cdot$cm). 
The wire is a fishing line with a diameter of 0.74 mm entrained by two lateral transmissions toothed belts and two stepping motors located at the top of the device.
The resulting wire velocity is $V = 100 $~cm/s. 
The width $w$ of the generated films is varied by a factor of two: we will study films with $w = 35$ and $70$~cm. 
The height $H$ of the films is in our study between 20 and 120~cm and four values are considered: $H = 20$, $40$, $80$, and $120$~cm. 

The soap film lifetime is recorded automatically by light reflectivity. 
A white light is shed at the bottom of the soap film and the reflection is measured by a detector composed of three photoresistors.
The soap film rupture leads to a loss of light detection. 
This automation allows performing robust statistical measurements of the lifetimes of the films, as we will see in the following.

We also measure the film thickness $h$ locally, using a UV-VIS spectrometer (Ocean Optics Nanocalc 2000) associated with an optical fiber of 400 $\mu$m diameter at different heights and at an equal distance between the two belts as shown in Fig.~\ref{fig:Setup}(a).

The humidity in the chamber of size  2.2~$\times$~1.0~$\times$~0.75~m$^3$ where the films are generated  is controlled with two commercial humidifiers, which allow having a homogeneous humidity in the chamber with a precision of 4 \%: one humidifier is placed at the top and the second is located at the bottom. The humidity is measured throughout the experiments with a SHT25 sensor (purchased from RS). Note that reaching atmospheric humidities higher than 80 \% is challenging for such a large box. We thus choose to work between 40 and 80 \%. This sensor also allows measuring the room temperature, which is about 22~$^{\circ}$C during all the experiments. 

\section{Results }

\subsection{Experimental measurements} 

\begin{figure}[!ht]
  \centering
    \includegraphics[width=\linewidth]{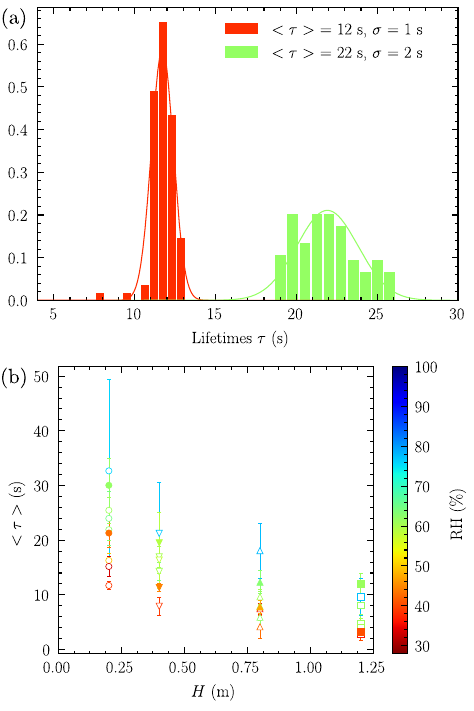}
  \caption{(a) Lifetime distributions measured for soap films of height $H = 20$~cm and width $w = 70$~cm, for different humidity conditions: $\text{RH} = 39 \ \%$ (in red) and $\text{RH} = 63 \ \%$ (in green). These histograms are obtained for about 100 films. The corresponding normal distributions are superimposed. 
  (b) Average lifetimes measured for different humidity conditions (indicated by the colormap). Two film widths are studied: $w = 35$~cm (full symbols)  and $w = 70$~cm (empty symbols). The error bars correspond to the standard deviation measured for each cycle studied of a hundred films. 
  }
  \label{fig:rawdata}
\end{figure}

In this section, we present the results of automated film lifetime measurements obtained from our device. To ensure statistical robustness, we systematically measure the lifetimes with cycles of 100 films for each probed height and humidity condition.

Two examples of histograms for films of height $H=20$~cm, corresponding to two distinct relative humidity values ($\text{RH} = 39 \ \%$ and $\text{RH} = 63 \ \%$) are given in Fig. \ref{fig:rawdata} (a). 
It can be observed that the distributions are well-defined for each humidity, and the solid lines represent the normal distributions fitted over the data. As expected, the average lifetime decreases as humidity decreases, varying on average from 22 seconds to 12 seconds in our example, when changing the relative humidity from 63 \% to 39 \%.

The dependence of the mean lifetime of the soap films on the height is given in Fig. \ref{fig:rawdata} (b). 
In this graph, two different widths are represented: $w = 35$~cm and $w = 70$~cm. 
The first tendency that we observe is that for each height probed, the lifetime decreases when $\text{RH}$ decreases. 
The other important observation is that the lifetime  $\tau$ decreases when $H$ increases for a given value of $\text{RH}$. We can also note a slight decrease of $\tau$ with $w$. 

As explained in the introduction, we propose that the film lifetime is mainly fixed by the thinning rate. 
This hypothesis is supported by the distributions shown in Fig. \ref{fig:rawdata} (a): the width of the distribution, which is due to the stochastic rupture of the films is indeed smaller than the variation of the mean value from one experiment to another.
Therefore, we focus our analysis on the thinning dynamics.

\subsection{Thinning of giant films}
\label{sec:ThinningOfGiantSopaFilms}%
\begin{figure}[!ht]
  \centering
    \includegraphics[width=\linewidth]{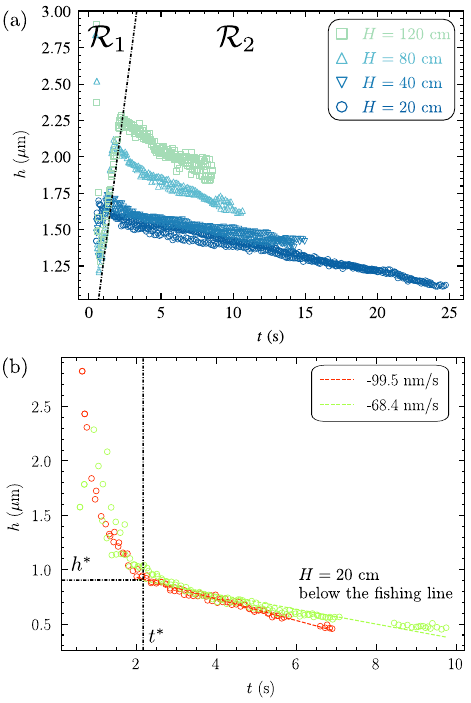}
  \caption{ Evolution of the thickness $h$ of the soap films during and after the generation at 1~m/s as a function of time ($w = 70$~cm). 
  (a) The measurements are performed 13~cm above the liquid bath (at the center of the films) at $\text{RH} = 60 \ \pm \ 4 \%$. 
  The dashed lines represent linear fits to measure the film thinning velocity, in the linear regime $\mathcal{R}_2$. 
  In (b), the measurements are performed below the fishing line (at the center of the films, $H = 20$~cm) at two humidity values: $\text{RH} = 39 \ \%$ (in red) and $\text{RH} = 63 \ \%$ (in green). The dashed line represents a linear fit to measure the film thinning velocity, in the linear regime $\mathcal{R}_2$. The slope is given in the plot.
}
  \label{fig:slope}
\end{figure}

\begin{figure}[!ht]
  \centering
    \includegraphics[width=\linewidth]{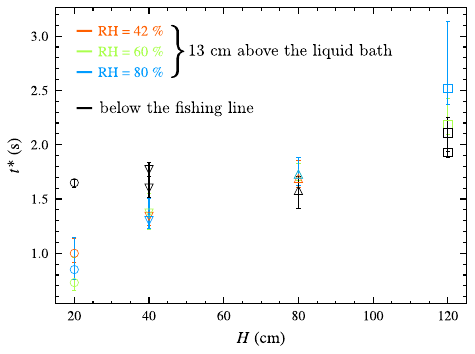}
  \caption{Evolution of the time $t^{\ast}$ as a function of the film height for different relative humidity values ($w = 70$~cm). The humidity values given in the figure have an accuracy of $\pm \ 4 \%$. The colored points correspond to measurements that are performed at 13~cm above the liquid bath (at the center of the films). The black points correspond to the measurements of $t^{\ast}$ using the data measured under the fishing line at $\text{RH} = 42 \ \%$ and/or $\text{RH} = 60 \ \%$. The symbols correspond to those used in Figs. \ref{fig:rawdata} and \ref{fig:slope}, with each film height represented by a different marker.
}
  \label{fig:t_star}
\end{figure}

Figure \ref{fig:slope} shows thinning curves measured respectively at the bottom of the films, 13 cm above the bath, for films of four different heights at $60$~\% humidity (Figure \ref{fig:slope}(a)), and at the top of the film below the fishing line of a 20~cm high static film, for two humidity values ($\text{RH} = 39 \ \%$ and $\text{RH} = 63 \ \%$) (Figure \ref{fig:slope}(b)). 

At the bottom of the film (Figure \ref{fig:slope}(a)), two regimes can be observed. 
During a first regime, $\mathcal{R}_1$, the film thickness increases before reaching a maximum and then, decreases almost linearly in time, which constitutes the second regime $\mathcal{R}_2$. 
The first regime has been the subject of a previous article \cite{pasquet2023thickness}, where we have drawn the following conclusions.
During and just after the generation, the film thickness profile is exponential, with a characteristic length fixed by the balance between gravity and the elasticity of the soap film.
The thickness evolution in this first regime, is linked to stress relaxation in the film following generation. The time $t^{\star}$, at which the second regime starts, corresponding to the maximum thickness, is plotted in color in Fig. \ref{fig:t_star}.

We observe systematically that these giant soap films burst from the top (see video in SI). 
We thus focus on the thinning dynamics at the top, in the vicinity of the fishing line (Figure \ref{fig:slope}(b)).
Here, the first regime $\mathcal{R}_1$ corresponds to quite high thinning rate, whereas a linear regime is also observed in the regime $\mathcal{R}_2$.
The transition time $t^*$, corresponding to the apparition of a linear regime, is plotted in black in Figure \ref{fig:t_star} and is very close to the one extracted from the thinning rate measured at the bottom of the film. 
Note that with the spectrometer, it is difficult to measure thicknesses below 400~nm, so that our last measured point in Figure \ref{fig:slope} does not necessarily correspond to the moment of film rupture.
The thinning velocity in the second regime is of the order of tens of nanometers per second and depends on the relative humidity and on the position in the film. 

\section{Lifetime of giant soap films}
The total lifetime $\tau$ of the films is the sum of the transition time $t^*$ and the lasting time of the second regime $t_{\mathcal{R}_2}$.
The value of $t^*$ is the measured value, extracted from the experimental data, since there is no theoretical prediction of its value. Fortunately, $t^*\le t_{\mathcal{R}_2}$ in many situations.
Thus, we will use the measured value of $t^*$ and develop a model to describe the thinning in the second regime, and to predict $t_{\mathcal{R}_2}$.

\subsection{A negligible drainage}
\label{sec:Thinning}

In absence of evaporation the downward Poiseuille flow in a foam film due to the gravitational field with the $z$-axis pointing downwards, writes \cite{Mysels1959}
\begin{equation} \label{eq:Poiseuille}
    \frac{\partial h}{\partial t} =  - \frac{ \rho_w g }{4 \eta} h^2 \frac{\partial h}{\partial z}, 
\end{equation}
where $\rho_w$ and $\eta$ are respectively the density and the dynamic viscosity in the film and $g$ the gravitational acceleration constant. 
We can note an important dependence on the thickness in $h^3$ in Eq. \eqref{eq:Poiseuille}. 
Nevertheless, the maximum film thickness observed in the regime $\mathcal{R}_2$ is around 1 $\mu$m. 
In this situation, a simple scaling analysis gives a thinning rate $\frac{\partial h}{\partial t} \simeq - \frac{\rho_w g}{4 \eta} h^3/H$. 
For the density and the viscosity of water and $H = 2$ m, the thinning rate is thus around $10^{-12}$ m/s, which is negligible compared to the actual thinning rate raging from a few tens to a hundred nanometers per second as shown in Figure~\ref{fig:slope}. 

This is why, as explained in the introduction, the drainage is attributed to marginal regeneration.
Nevertheless, in these giant soap films, we never observed any development of marginal regeneration at the bottom meniscus.
Additionally, the lateral gutters, close to the meniscus, which are crucial in centimetric films, are not visible, and their contribution is probably negligible. 
Indeed, if they exist, their width is much smaller than the total film width in contrast to centimetric films. 
Thus, we can rule out the drainage due to marginal regeneration.

We finally propose to attribute the thinning in the linear regime $\mathcal{R}_2$ to evaporation, in good agreement with the order of magnitude of the thinning rate, about tens of nanometers per second \cite{BoulogneDollet2018}. In the next section, we will estimate a
value for $\tau_{\mathcal{R}_2}$ within these hypotheses.

\subsection{Prediction of the evaporation timescale}
\begin{figure}[!ht]
  \centering
    \includegraphics[width=\linewidth]{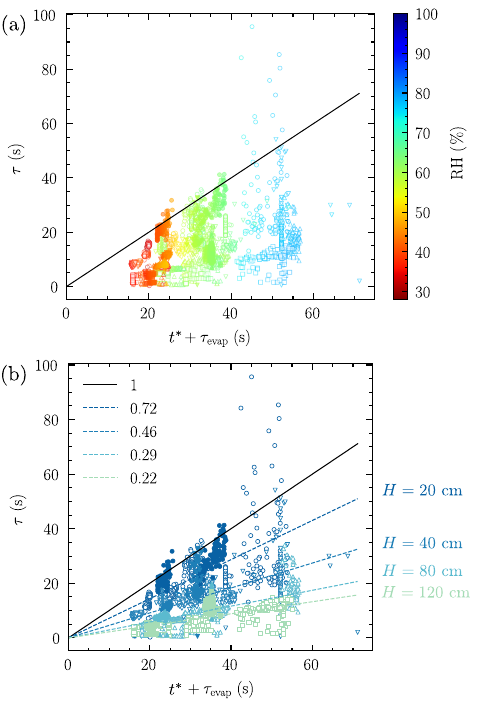}
  \caption{(a) Evolution of the measured film lifetime $\tau$ versus the lifetime predicted by the Eq. \eqref{eq:def_tau_max}. The colors correspond to the humidity $\text{RH}$ and are indicated using the vertical color bar. 
  Four static film heights are studied ($H = 20$~cm, $H = 40$~cm, $H = 80$~cm and $H = 120$~cm) as well as two distinct film widths: $w_0 = 35$~cm (full symbols) and $w_0 = 70$~cm (empty symbols). 
  The black line corresponds to the equality between the axis. 
  (b) Identical graph with different colors, corresponding to the height of the films $H$. 
  The dotted lines correspond to a linear fit performed for all the measured lifetimes for films of a given height $H$. 
  The slopes are indicated in the legend.}
  \label{fig:alllifetimes}
\end{figure}
To estimate the lifetime of the films as a function of the humidity conditions present in the chamber, we can estimate their evaporation rate. 
This necessitates modeling the concentration gradient of vapor established from the surface of the evaporating film to its equilibrium value, far from the interface.
For water-based evaporating surfaces larger than a typical scale of about one centimeter, the convection due to the lower density of humid air compared to dry air must be considered \cite{Shahidzadeh-Bonn2006,Dollet2017,BoulogneDollet2018}. 
The transition toward a convective evaporation is quantified by the Grashof number 
\begin{equation}  
\label{eq:Grashof}
\text{Gr} \  = \ \dfrac{\mid \Delta \rho_{\text{a}} \mid}{\rho_{\text{a}},\infty} \ \dfrac{g}{\nu^2} \ \mathcal{L}^{3},
\end{equation}
\noindent where $\nu$ is the kinematic viscosity of air ($\nu = 1.55 \ 10^{-5}$~m$^2$/s at 20~$^{\circ}$C),
$\rho_{\text{a},\infty}$ is the density of the ambient air (away from the evaporating interface), 
$\mid \Delta \rho_{\text{a}} \mid = \mid \rho_{\text{a, sat}} - \rho_{\text{a},\infty} \mid$ is the difference in density between the air at the surface and the ambient air, 
and $\mathcal{L}$ is the vertical characteristic length over which the concentration gradient of vapor is established.
Natural convection must be taken into account when $\text{Gr} \gg 1$. 
A first determination of the Grashof numbers, using the Eq. \eqref{eq:Grashof} for an intermediate humidity corresponding to $\text{RH} = 60 \ \%$, indicates that for our problem: $\text{Gr} > 10^{6}$. 
The convection flux due to the evaporation of these large vertical films must therefore be taken into account.

We propose to use the calculation of the natural convection flow on a vertical plate developed by Schmidt and Beckmann \cite{Schmidt1930} adapted for the evaporation of hydrogel slabs up to 20 cm high \cite{BoulogneDollet2018}. 
Their calculation of the local evaporative flux for a vertical film, which assumes a laminar boundary layer, leads to:
\begin{equation} \label{eq:BoulogneDollet}
j_{\text{evap}} \ \approx \ \widehat{c_0} \ \dfrac{D \left( c_{\text{sat}} - c_{\infty}   \right)}{\rho_w}   \ \dfrac{\text{Gr}^{1/4}}{\mathcal{L}},
\end{equation}
\noindent where $D$ is the water vapor diffusion coefficient in air ($D = 2.44 \ 10^{-5}$~m$^2$/s at 20~$^{\circ}$C), $\rho_{\text{w}}$ is the density of water, $c_{\text{sat}}$ and $c_{\infty}$ are respectively the mass concentrations of saturated vapor and water vapor in the air ($c_{\text{sat}} = 17$~g/m$^3$ at 20~$^{\circ}$C), and $\widehat{c_0}$ is a numerically determined prefactor ($\widehat{c_0} \simeq 0.478$). 
This model is developed for centimetric vertical films made of hydrogel, for which water evaporates in the atmosphere.
The characteristic size $\mathcal{L}$ on which the boundary layer is developed corresponds to the height $H$ of the film.  

Thus, the thinning rate due to evaporation writes
\begin{equation} \label{eq:dhdt}
\frac{\partial h}{\partial t} \ = \ 2 \ j_\text{evap},
\end{equation}
where the factor of two takes into account the contribution of both liquid-air interfaces of the films.
As $j_{\rm evap}$ remains constant over time, this equation is in agreement with the linear decrease of the film thickness observed experimentally, which tend to confirm that evaporation is the main thinning mechanism in this configuration.

Eq. \eqref{eq:dhdt} directly leads to a prediction for the evaporation time:
\begin{equation} \label{eq:def_tau_max}
\tau_{\text{evap}} \ = \ \dfrac{h^{\ast} \ - \ h_{\text{rupt}}}{2 \ j_\text{evap}}
\end{equation}
\noindent
with $h_{\text{rupt}}$ the thickness at which the films break. 
We consider here the thickness at the top of the films, since this is where $h$ is minimum and thus the films are more likely to rupture. 

In the next section, we will compare the lifetime of the films with $t^* + t_{\mathcal{R}_2}$ with $t^*$ measured experimentally (Figure \ref{fig:t_star}) and $t_{\mathcal{R}_2} = \tau_{\text{evap}}$.

\subsection{Discussion}
In Fig. \ref{fig:alllifetimes} (a), we plot all the measurements made for different heights and humidity conditions as a function of the lifetime given by Eq. \eqref{eq:def_tau_max}, in the case where $h_{\text{rupt}} = 0$. 
This is the maximum time that can be predicted for $\tau_{\text{evap}}$, considering that the films break when their thickness is equal to zero. 
On this figure, several thousands of films are represented and the solid line corresponds to a line of equation $\tau = t^* + \tau_{\rm evap}$. 
This graph shows that the vast majority of the measured lifetimes are below this limit.

This graph is shown again in Fig. \ref{fig:alllifetimes} (b), with a different color scale, which corresponds to the height of the films $H$. 
For the films of height $H = 20$ cm, we observed that the measured lifetimes are below but close to the theoretical prediction (solid black line). 
This suggests that the model of Boulogne \& Dollet \cite{BoulogneDollet2018} can describe the evaporation time of these soap films as well. 
In this case, the discrepancy between the prediction and measurements is attributed to the rupture of the films when they are still at non-zero thicknesses. 

As the height of the films increases, the films tend to break systematically at smaller time. 
For each film height probed, a dotted line shows the linear fit of all lifetimes as a function of $t^{\star} + \tau_{\text{evap}}$.
The measured slope is given in the legend of the plot in figure \ref{fig:alllifetimes} (b). 
We notice that as the height of the films increases, the value of this slope decreases and deviates from unity. 
This suggests that the main physical ingredients of the model are still holding, but that the evaluation of the evaporation rate is no longer suitable for higher films. 
The best candidate to explain this mismatch is the value of the film length $\mathcal{L}$, which we took equal to $H$ following the argument presented in reference \cite{BoulogneDollet2018}.
Nevertheless, this length seems to be smaller as the film height increases.
As suggested by Haaland and Sparrow, the boundary layer can undergo a destabilization due to the difference of buoyancy with the surrounding atmosphere that causes a more complex three-dimensional flow increasing the evaporation of the film \cite{Haaland1973}.
Such a mechanism, that must be confirmed with dedicated experimental investigations, would explain the difference between the prediction of the model and the measured film thinning.

\section{Conclusion}
In this paper, we have presented a study of the stability of soap films having a height between 20~cm and 1.20~m, generated at a fixed velocity of 1.0~m/s in a humidity-controlled chamber. 
We have highlighted the influence of their width and of the relative humidity on their stability: as expected and already observed on other small soap systems \cite{miguet2020stability}, as the relative humidity $\text{RH}$ increases, the lifetime of these large static films is also increased. 

To quantify this observation, we propose that evaporation is the predominant mechanism.
The convection must be considered in the description of the evaporation of our films, since the Grashof numbers are considerable. 
We therefore compared the lifetimes measured to the evaporation times predicted by Boulogne \& Dollet \cite{BoulogneDollet2018}. 
This gives a good prediction of the maximum value of the lifetimes for all film sizes and a quantitative prediction for the 20 cm films.
Nevertheless, a full description of the films lifetime would necessitate a comprehensive understanding of the first regime and thus of film relaxation as well as a better understanding of the evaporation of vertical liquid sheets at a meter scale. 
In particular, the regime at high humidity, above 80 \%, identified by Frazier \textit{et al} \cite{Frazier2020} would certainly be very interesting to explore.
In addition, the recent findings on the cooling of soap films induced by evaporation raises also the question of the significance of the thermal effects in the convective regime, which must be addressed in future studies \cite{Boulogne2022,Boulogne2023,Corpart2024}.

\section*{Acknowledgments} 
Funding from ESA (MAP Soft Matter Dynamics), CNES (through the GDR MFA) and the ANR DRAINFILM is acknowledged. We thank Isabelle Cantat for fruitful discussions and Sandrine Mariot for the design and development of the experimental setup.

\bibliography{Biblio} 
\bibliographystyle{unsrt}

\end{document}